\documentstyle[epsfig]{aipproc}

\newcommand{\ra}{\rightarrow}
\renewcommand{\a}{\alpha}
\newcommand{\s}{\hat s}
\newcommand{\th}{\hat t}
\newcommand{\nn}{\nonumber}
\newcommand{\beq}{\begin{equation}}
\newcommand{\eeq}{\end{equation}}

\begin{document}
\title{BFKL: a minireview\thanks{Invited talk at DIS97}}

\hspace*{\fill}\parbox[t]{4cm}{EDINBURGH 97/8\\ June 1997}

\author{Vittorio Del Duca}
\address{Particle Physics Theory Group,\,
Dept. of Physics and Astronomy\\ University of Edinburgh,\,
Edinburgh EH9 3JZ, Scotland, UK}
\maketitle

\begin{abstract}
\noindent
We summarize the state of the art of resummation techniques
in the small-$x$ limit of perturbative QCD.
\end{abstract}

\section*{The BFKL resummation}
Hard strong-interaction processes are characterised by a large
scale, e.g. the center-of-mass energy in $e^+e^-\ra q\bar q$,
the dilepton invariant mass in Drell-Yan production, the momentum
transfer $Q$ in DIS, the jet transverse energy in hadronic
jet production. Calculations are performed at a fixed order in the
coupling constant $\a_s$. Large logarithms of the ratio
of the renormalization scale to the scales above appear, but
the renormalization group lets us resum them appropriately. 
Namely, we may organise the terms in the perturbative expansion
in $\a_s$ in order to resum the large logarithms by letting the
coupling constant run. However, there are processes characterised by 
two large and disparate scales, like DIS at small $x$, with $x=Q^2/s$
the squared ratio of the momentum transfer to the lepton-hadron 
center-of-mass energy, or hadronic dijet production at large rapidity 
intervals $\Delta y$, with $\Delta y\simeq\ln(\s/|\th|)$,
$\s$ the squared parton center-of-mass energy and $|\th|$ of the
order of the squared jet transverse energy, where also the logarithms
$\ln(1/x)$ or $\ln(\s/|\th|)$ are large, and may have to be resummed. 
This is topical nowadays because of the large kinematic region 
explored in DIS by HERA, where values of $x$ of the order of $10^{-5}$
have been attained. The Balitsky-Fadin-Kuraev-Lipatov (BFKL) equation 
\cite{bal} lets us resum these logarithms.

\section*{The QCD amplitudes in the $\s\gg|\th|$ limit}

Let us consider parton-parton scattering as a paradigm process.
For $\s\gg|\th|$ it is dominated by gluon exchange in the
$\th$ channel, thus the functional form of the amplitudes
for gluon-gluon, gluon-quark or quark-quark scattering is the
same; they differ only for the color strength in the parton-production
vertices. For example the amplitude for $g\,g\to g\,g$ scattering, 
with all the external gluons outgoing, may be written in a helicity
basis as \cite{FKL}, \cite{ptlip}
\begin{equation}
M^{aa'bb'\,{\rm tree}}_{\nu_a\nu_{a'}\nu_{b'}\nu_b} = 2 \s
\left[i g\, f^{aa'c}\, C^{gg(0)}_{-\nu_a\nu_{a'}}(-p_a,p_{a'}) \right]
{1\over\th} \left[i g\, f^{bb'c}\, C^{gg(0)}_{-\nu_b\nu_{b'}}(-p_b,p_{b'}) 
\right]\, ,\label{elas}
\end{equation}
with $\nu$ the helicity of the external gluons, and $C^{gg(0)}$ the 
helicity-conserving vertex $g^*\, g \rightarrow g$, 
with $g^*$ an off-shell gluon, 
\begin{equation}
C_{-+}^{gg(0)}(-p_a,p_{a'}) = -1\qquad C_{-+}^{gg(0)}
(-p_b,p_{b'}) = - {p_{b'\perp}^* \over p_{b'\perp}}\, ,\label{centrc}
\end{equation}
and $p_{\perp}=p_x+ip_y$ the complex transverse momentum. 
The vertices transform into their 
complex conjugates under helicity reversal.
For gluon-quark or quark-quark scattering, we only need to exchange
the structure constants with color matrices in the fundamental representation
and change the vertices $C^{gg(0)}$ to $C^{\bar q q(0)}$ \cite{thuile}.
Next, we compute the $O(\a_s)$ corrections. In order to do that, we
must consider the emission of an additional gluon, i.e. the amplitude
for the production of three gluons, which are taken to be strongly
ordered in rapidity, $y_{a'} \gg y \gg y_{b'}$ and with comparable 
transverse momenta $|p_{a'\perp}|\simeq|k_\perp|\simeq|p_{b'\perp}|$.
This is the simplest example of {\sl multi-Regge kinematics}.
The scattering amplitude is then,
\begin{eqnarray}
M^{tree}_{g g \ra g g g} &=& 
2 {\hat s} \left[i g\, f^{aa'c}\, C_{-\nu_a\nu_{a'}}^{gg}(-p_a,p_{a'})
\right]\, {1\over\hat t_1}\, \label{three}\\ &\times& \left[i g\,f^{cdc'}\, 
C^g_{\nu}(q_1,q_2)\right]\, {1\over \hat t_2}\, 
\left[i g\, f^{bb'c'}\, C_{-\nu_b\nu_{b'}}^{gg}(-p_b,p_{b'}) \right]\, ,\nn
\end{eqnarray}
with $p_{a'\perp} = - q_{1\perp}$, $p_{b'\perp} = q_{2\perp}$ and
$\th_i \simeq - |q_{i\perp}|^2$ with $i=1,2$ and with the Lipatov vertex
$g^*\, g^* \rightarrow g$ \cite{lip}, \cite{lipat}, \cite{ptlip},
\beq
C^g_+(q_1,q_2) = \sqrt{2}\, {q^*_{1\perp} q_{2\perp}\over k_\perp}\, 
.\label{lipeq}
\eeq
The amplitude (\ref{three}) has the effective form of a gluon-ladder 
exchange in the $t$ channel, however the additional gluon $k$ has been 
inserted either along the ladder or as a 
bremsstrahlung gluon on the external legs. In this sense the Lipatov
vertex (\ref{lipeq}) is a non-local effective vertex. Eq.(\ref{three})
generalizes to the production of $n+2$ gluons \cite{FKL}, \cite{ptlip} 
in the multi-Regge kinematics, $y_{a'} \gg y_1 \gg ...\gg y_n \gg y_{b'}$,
with $|p_{i\perp}|\simeq|p_{\perp}|$ and $i=a',1,...,n,b'$, 
in a straightforward manner. The square of the amplitude (\ref{three}), 
integrated over the phase space of the intermediate gluon in
multi-Regge kinematics yields an $O(\a_s\ln(\s/|\th|))$ correction
to gluon-gluon scattering, which is however infrared divergent.
To complete the $O(\a_s)$ corrections, and cancel the infrared
divergences, we must compute the 1-loop gluon-gluon amplitude
in the leading logarithmic (LL) approximation in $\ln(\s/|\th|)$.
It turns out that in the LL approximation the virtual corrections
to the parton-parton scattering amplitude may be computed to all
orders in $\a_s$ \cite{FKL}, \cite{zvi} by simply replacing in the gluon
propagator of eq.~(\ref{elas})
\begin{equation}
{1\over\th} \to {1\over\th} 
\left({\s\over -\th}\right)^{\alpha(\th)}\, ,\label{sud}
\end{equation}
where $\alpha(\th)$ is related to the loop transverse-momentum integration
\begin{equation}
\alpha(\th) \equiv g^2 \alpha^{(1)}(\th) = \alpha_s\, N_c\, \th \int 
{d^2k_{\perp}\over (2\pi)^2}\, {1\over k_{\perp}^2
(q-k)_{\perp}^2}\qquad \th = q^2 \simeq - q_{\perp}^2\, .\label{allv}
\end{equation}
It is conventionally said that because of eq.~(\ref{sud}) the gluon
is {\sl reggeised}. Adding the 1-loop gluon-gluon amplitude,
multiplied by its tree-level counterpart,
to the square of the amplitude (\ref{three}), 
integrated over the phase space of the intermediate gluon, cancels
the infrared divergences and yields a finite $O(\a_s\ln(\s/|\th|))$ 
correction to gluon-gluon scattering. Eq.~(\ref{sud}) is conjectured
to generalize
to the production of $n+2$ gluons in the multi-Regge kinematics,
namely the LL virtual corrections to all orders in $\a_s$ to the tree
level amplitude in the multi-Regge kinematics are obtained by
the replacement (\ref{sud}) for each $\th_i$ \cite{FKL}.
This may be checked at the one loop level, at least for
the Parke-Taylor helicity configuration, by taking the one-loop $N=4$
supersymmetric amplitudes \cite{bdk}. This is due to the fact that
in the LL approximation the only particle circulating in a loop in an
$N=4$ supersymmetric amplitude as well as in a regular QCD amplitude
is the gluon. Therefore in the LL approximation $N=4$
supersymmetric amplitudes and QCD amplitudes coincide.
Since the LL terms are universal, i.e. they are the same no matter
what kind of partons initiate the scattering, they may be thought of
as radiative corrections to a gluon propagator exchanged in the $\th$
channel. Squaring then the amplitude for the production of $n+2$ 
gluons in the multi-Regge kinematics, integrating over the 
phase space and summing over $n$, we obtain a resummation of the terms
of $O(\a_s^n\ln^n(\s/|\th|))$ embodied in the BFKL equation, which
describes the evolution of the gluon propagator in transverse momentum
in the $\th$ channel.

There are, though, a few drawbacks in the BFKL formalism. First, 
energy and longitudinal momentum are not conserved; the terms that violate
energy-momentum conservation are formally subleading, however they may
be important for any phenomenological use of the BFKL equation. A way
to avoid that is to obtain, through a Monte Carlo event generator, a 
numerical solution of the BFKL equation that exhibits energy-momentum 
conservation \cite{carl}. Secondly, the BFKL equation has been
derived at fixed coupling constant, thus any variation in the scale
of the coupling constant, $\a_s(\nu^2)=\a_s(\mu^2) - 
b_0\ln(\nu^2/\mu^2)\a_s^2(\mu^2) + \dots$, with $b_0= (33-2n_f)/
12\pi$ and $n_f$ the number of quark flavors, would appear in the 
next-to-leading-logarithmic (NLL) terms, because it yields terms
of $O(\a_s^n\ln(\nu^2/\mu^2)\ln^{n-1}(\s/|\th|))$.
However, scale variations of $\a_s$
are phenomelogically important. Thirdly, because of the strong
rapidity ordering any two-parton invariant mass is large. Thus there
are no collinear divergences, and jets are determined
only to leading order and accordingly have no non-trivial structure.

\section*{The next-to-leading-logarithmic corrections}

A way to alleviate these problems is to compute the 
NLL corrections to the BFKL equation.
The NLL corrections to the FKL amplitudes are divided into
real corrections, induced by the corrections to the multi-Regge kinematics,
and virtual NLL corrections.
The real corrections to the tree-level FKL amplitudes arise from the 
kinematical regions in which two partons are produced with
similar rapidity, either at the ends of or along the ladder.
The building blocks of these amplitudes are the vertices which
describe the emission of two partons in the forward-rapidity region, 
$g^*\, g \rightarrow g\, g$ \cite{fl}, \cite{ptlipnl}, 
$g^*\, g \rightarrow \bar{q}\, q$, \cite{fk}, \cite{ptlipqq},
and $g^*\, q \rightarrow g\, q$ \cite{thuile}; and 
in the central-rapidity region, $g^*\, g^* \rightarrow g\, g$ 
\cite{fl}, \cite{ptlipnl}, \cite{fl2}, and $g^*\, g^* \rightarrow 
\bar{q}\, q$, \cite{ptlipqq}, \cite{fl2}.

The calculation of the NLL corrections to the BFKL kernel is then
articulated as follows:
the vertices for the emission of two gluons or of a $\bar{q}\, q$
pair in the central-rapidity region, squared
and integrated over the phase space \cite{fl2}, \cite{ffkint} and
the one-loop correction to the Lipatov vertex \cite{flvirt},
\cite{ffq}, multiplied by its tree-level counterpart (\ref{lipeq}),
yield the real NLL corrections to the BFKL equation, while the virtual
ones are given by the NLL corrections \cite{ffk} to the gluon 
reggeisation (\ref{sud}).

I shall briefly describe how to compute the virtual corrections beyond
the LL approximation. In order to do that,
we need the general form of a scattering amplitude in the high-energy
limit. For gluon-gluon scattering (\ref{elas}), it is \cite{flvirt}
\begin{eqnarray}
M^{aa'bb'}_{\nu_a\nu_{a'}\nu_{b'}\nu_b} &=& s
\left[i g\, f^{aa'c}\, C^{gg}_{-\nu_a\nu_{a'}}(-p_a,p_{a'}) \right]
{1\over t} \left[\left({s\over -t}\right)^{\alpha(t)} +
\left({-s\over -t}\right)^{\alpha(t)} \right] \nonumber\\ &\times&
\left[i g\, f^{bb'c}\, C^{gg}_{-\nu_b\nu_{b'}}(-p_b,p_{b'}) 
\right]\, ,\label{elasb}
\end{eqnarray}
where now 
\begin{equation}
\alpha(t) = g^2 \alpha^{(1)}(t) + g^4 \alpha^{(2)}(t) + O(g^6)\,
,\label{alphb}
\end{equation}
and
\begin{equation}
C^{gg} = C^{gg(0)} + g^2 C^{gg(1)} + O(g^4)\, .\label{fullv}
\end{equation}
In the NLL approximation we need to know $C^{gg(1)}$, obtained by
expanding eq.~(\ref{elasb}) to $O(g^4)$, i.e. to one loop
\cite{flvirt}, \cite{ff} (accordingly by expanding the analog
of eq.~(\ref{elasb}) for quark-quark scattering, one gets
$C^{q\bar q(1)}$ \cite{ff}), and we need to know $\alpha^{(2)}(t)$,
obtained by expanding eq.~(\ref{elasb}) to $O(g^6)$, i.e. to two loops
\cite{ffk}. The one-loop corrections $C^{gg(1)}$ ($C^{q\bar q(1)}$)
to the $g^*\, g \rightarrow g$ ($g^*\, q \rightarrow q$) vertices 
(\ref{centrc}) do not enter
directly the NLL corrections to the BFKL kernel, however they are
necessary in order to correctly extract $\alpha^{(2)}(t)$ and
the one-loop correction to the Lipatov vertex; together with the
vertices for the emission in the forward-rapidity region, 
$g^*\, g \rightarrow g\, g$, $g^*\, g \rightarrow \bar{q}\, q$, 
and $g^*\, q \rightarrow g\, q$, they
are also the building blocks for the calculation of the NLO off-shell
coefficient functions for jet production.
Then $\alpha^{(2)}(t)$ yields the
 reggeisation of the
gluon at NLL accuracy, mentioned above.

Analogously, in order to compute the one-loop correction to the 
Lipatov vertex we need the general form of the amplitude \cite{flvirt}
for the production of three gluons (\ref{three}) beyond the
LL approximation (\ref{sud}) (a more detailed description may be found in 
ref.~\cite{fadin}).

The full NLL corrections to the kernel of the BFKL equation are not
available yet, as of now \cite{fadin}, however, the contribution
proportional to the number of quark flavors $n_f$, obtained by
putting together the square of the vertex for the emission of
a $q\bar q$ pair in the central-rapidity region, the quark-loop
corrections to the Lipatov vertex and the quark-loop contribution
to the NLL corrections to the reggeisation of the gluon, has been
obtained in ref.~\cite{cc}. When the full NLL corrections to the 
BFKL equation are known it will be possible to extract the NLL
contribution to the gluon anomalous dimensions $\gamma_{gg}$ and
$\gamma_{gq}$, the ones for the quark anomalous dimensions
$\gamma_{qg}$ and (the flavor singlet) $\gamma_{qq}$ being already
known \cite{ch}. Hence, provided the appropriate NLO off-shell
coefficient functions are computed, one will be able to obtain
the NLL corrections to DIS at small $x$ and to hadronic dijet 
production at large rapidity intervals.

\bigskip
I should like to thank Stefano Catani and Stefano Forte for
useful discussions, and the organizers of DIS97 for the support.

\end{document}